\documentclass[11pt,draftcls,onecolumn]{IEEEtran}
\usepackage[cmex10]{amsmath}
\usepackage{array}

\usepackage{amssymb}
\usepackage{amsmath}
\usepackage{amsthm}
\usepackage[lined,boxed,commentsnumbered, ruled]{algorithm2e}
\usepackage{xcolor} %改变字体颜色时要用

\usepackage{graphicx}%为了插入eps格式的图片
\usepackage{epstopdf}

\newcommand{\ua}{\uparrow}
\newcommand{\nc}{\newcommand}
\nc{\da}{\downarrow} \nc{\hc}{\hat{c}} \nc{\hS}{\hat{S}}
\nc{\bra}{\langle} \nc{\ket}{\rangle} \nc{\eq}{equation (\ref}
\nc{\h}{\hat} \nc{\hT}{\h{T}}\nc{\be}{\begin{eqnarray}}
\nc{\ee}{\end{eqnarray}}\nc{\rd}{\textrm{d}}\nc{\e}{eqnarray}\nc{\hR}{\hat{R}}\nc{\Tr}{\mathrm{Tr}}
\nc{\tS}{\tilde{S}}\nc{\tr}{\mathrm{tr}}\nc{\8}{\infty}\nc{\lgs}{\bra\ua,\phi|}\nc{\rgs}{|\ua,\phi\ket}
\nc{\hU}{\hat{U}}\nc{\lfs}{\bra\phi|}\nc{\rfs}{|\phi\ket}\nc{\hZ}{\hat{Z}}\nc{\hd}{\hat{d}}\nc{\mD}{\mathcal{D}}
\nc{\bd}{\bar{d}}\nc{\bc}{\bar{c}}\nc{\mc}{\mathcal}\nc{\ea}{eqnarray}\nc{\mG}{\mathcal{G}}\nc{\bce}{\begin{center}}
\nc{\ece}{\end{center}}
\ifCLASSOPTIONcompsoc
  \usepackage[caption=false,font=normalsize,labelfont=sf,textfont=sf]{subfig}
\else
  \usepackage[caption=false,font=footnotesize]{subfig}
\fi

\begin{document}

\title{The Unambiguous Distance in a Phase-based Ranging System with Hopping Frequencies}

\author{Yue Zhang, Wangdong Qi and Su Zhang

\thanks{The authors are with the PLA University of Science and Technology, Nanjing, Jiangsu 210007, China (e-mail: zhyemf@gmail.com; wangdongqi@ gmail.com; franklinzhang1985@gmail.com).}
}

% The paper headers
\markboth{submitted to iet electronics letters,~Vol.~X, No.~X, X~2014}%
{yue zhang \MakeLowercase{\textit{et al.}}: Performance of ML Range Estimator in Radio Interferometric Positioning Systems}

\maketitle

\begin{abstract}
It is a challenge to specify unambiguous distance (UD) in a phase-based ranging system with hopping frequencies (PRSHF). In this letter, we propose to characterize the UD in a PRSHF by the probability that it takes on its maximum value. We obtain a very simple and elegant expression of the probability with growth estimation techniques from analytic number theory. It is revealed that the UD in a PRSHF usually takes on the maximum value with as few as 10 frequencies in measurement, almost independent of the specific distribution of available bandwidth. 
\end{abstract}

\section{Introduction}
A phase-based ranging system with hopping frequencies (PRSHF) is a novel ranging system where phase shifts of carriers are measured to estimate the distance between a transmitter and a receiver. In contrast to traditional phase-based ranging systems such as OMEGA \cite{1}, carrier frequencies employed in PRSHF are chosen randomly from a set of frequencies \cite{2}. Since PRSHF can take advantage of the dynamic and possibly discontinuous available bandwidth provided by cognitive radio technology \cite{3}, it has great potential in anti-jamming localization and navigation.

The unambiguous distance (UD) is a crucial metric to gauge the performance of a phase-based ranging system in scalability. While it is straightforward to determine it in traditional phase-based ranging systems\cite{4}, it is a challenge to specify it in a PRSHF due to the randomness of frequencies and discontinuity of available bandwidth.

In this letter, we propose to characterize the UD in a PRSHF by the probability that it takes on its maximum value which is equivalent to the probability that integers randomly chosen from a specific set are relatively prime. With growth estimate techniques from analytic number theory, we show that the above probability depends on the number of frequencies employed in measurement in a simple and elegant way. In fact, the probability of UD takes on the maximum value is close to 1 when a PRSHF employs as few as 10 frequencies in measurement, almost independent of the specific distribution of the available bandwidth.

\section{System model of a PRSHF}
In a phase-based ranging system, a sender transmits signals $x(t)=A_{t}\cos(2\pi f_{i}t + \varphi_{0})$and a receiver gets signals $y(t)=A_{r}\cos(2\pi f_{i}(t-R/c) + \varphi_{0})$ where $f_{i}(i=1,\cdots ,M)$ are the carrier frequencies in the available bandwidth of radio transceivers, c is the speed of signal propagation and $R$ is the distance between the sender and the receiver. The phase shifts between transmitted and received signals
\begin{align}\label{phi}
  \varphi_{i}=2\pi\frac{f_{i}}{c}R \left(mod\ 2\pi\right),i=1,\cdots ,M
\end{align}
are measured and used in the resolution of distance $R$.

In a PRSHF, $M$ frequencies employed in measurement are chosen randomly from the $N=\sum_{l=1}^L N_{l}$ available frequencies consisting of $L$ possibly separated segments with $N_{l}$ frequencies in the $l$th segment

We assume that all frequencies in the bandwidth of the transceivers are multiples of the minimum frequency interval $f_{min}$ with $n_{0}f_{min}$ and $(n_{0}+N_{0}-1)f_{min}$ denote the smallest and largest of them. Obviously, the total number $N_0$ of frequencies is at least as large as the number $N$ of available frequencies. We assume that $N$ is on the same order as $N_0$.

\section{UD in a PRSHF}
It can be easily verified that UD in the phase-based ranging system with a fixed set of $M$ frequencies $\{f_{i}=k_{i}f_{min}\}$ in measurement is
\begin{align}\label{UD}
  UD=\frac{c}{kf_{min}}
  %UD=c/(kf_{min})
\end{align}
where $k$ is the greatest common divisor (GCD) of $\{k_{i}\}$, if we consider the fact that the UD is the least common multiple (LCM) of $\{\lambda_{i}\}$, where $\lambda_{i}=c/f_{i}=c/(k_if_{min})$  is the wavelength of the $i$th carrier \cite{4}.

In a PRSHF, since the set of carrier frequencies is not fixed but chosen randomly from the available frequencies, the UD is essentially a random variable. Considering that $c/f_{min}$, the maximum value of UD, is large enough to accommodate most scenarios because the minimum frequency interval $f_{min}$ is very small (well below 1Hz in some modern transceivers \cite{5}), we propose to characterize the UD not by its entire distribution but by the probability that UD takes on the maximum value.

Let $\mathbb{N}$ denote the set of positive integers $n$ where $nf_{min}$ is in the set of available frequencies. Obviously, $\mathbb{N}$ consists of $L$ segments of positive integers with $N$ elements. It can be seen that UD takes on the maximum value in a PRSHF when $M$ integers $\{k_{i}\}$ chosen randomly from $\mathbb{N}$ are relatively prime. The calculation of the probability, however, is very difficult since it depends not only on $M$, but on $N$, $L$ and the set $\mathbb{N}$ which is determined by the distribution of the available bandwidth.

As it turns out, if we consider the practical constraint that $L$ and $M$ are much smaller than $N$ and drop the hope of finding the exact expression of the probability but focus on its ``growth'' with respect to those parameters, we can specify it rather well.

Growth estimation techniques used in the following are standard in analytic number theory. Readers are referred to \cite{6,7} for a look of the simpler case when $L$ is 1 and $n_{0}$ is 1. We focus here on difficulties arising from the arbitrary starting point $n_{0}$ and arbitrary distribution of the available bandwidth in our system model.

let $p_{1},p_{2},\cdots$ denote distinct primes. To simplify the analysis, we assume $M$ is greater than 2. The following notations will be used:

$A_{p_{1}\cdots p_{n}}$: Number of $M$-tuple of positive integers belonging to $\mathbb{N}$ which can be divided by $\prod_{i=1}^{n}p_{i}$.

$Z$: Number of $M$-tuple of positive integers belonging to $\mathbb{N}$ which are relatively prime.

$P$: Probability that $M$-tuple of positive integers chosen at random from $\mathbb{N}$ are relatively prime.

Obviously, $P=Z/N^M$. Since $M$-tuple of integers are relatively prime if and only if there exists no prime that divides all $M$ integers, the Inclusion-Exclusion Principle shows that
\begin{align}\label{zmn}
  Z=N^M-\sum_{p_1}A_{p_1}+\sum_{p_1<p_2}A_{p_1p_2}-\sum_{p_1<p_2<p_3}A_{p_1p_2p_3}+\cdots
\end{align}
By using the M\"{o}bius function $\mu$ \big(M\"{o}bius function is a number theoretic function defined as: $\mu(d)=0$ if $d$ has one or more repeated prime factors; $\mu(d)=1$ if $d=1$; $\mu(d)=(-1)^r$ if $d$  is a product of $r$ distinct primes), we rewrite (\ref{zmn}\big) more compactly as

\begin{align}\label{zmn2}
  Z=\sum_{j=1}^{\infty}\mu(j){x_j}^M
\end{align}
where $x_j$ denotes the number of integers in $\mathbb{N}$ which can be divided by $j$. It can be easily verified that ${x_j}^M$ is the number of $M$-tuple of positive integers in $\mathbb{N}$  which can be divided by $j$. If we define $Z_1=\sum_{j=1}^{N_0}\mu(j){x_j}^M$ and $Z_2=\sum_{j=N_0+1}^{\infty}\mu(j){x_j}^M$ , then $Z=Z_1+Z_2$.

Next we try to find $Z_1$. Let $x_j(l)$ denote the number of integers belonging to the $l$-th segment of $\mathbb{N}$ which can be divided by $j$, then $x_j=\sum_{l=1}^L x_j(l)$. Apparently $x_{j}(l)=\left\lfloor \frac{N_l}{j}\right\rfloor or \left\lfloor\frac{N_l}{j}\right\rfloor+1$, so we have $\sum_{l=1}^L (\frac{N_l}{j}-1)\leq x_j \leq\sum_{l=1}^L (\frac{N_l}{j}+1)$, and then $\tfrac{N}{j}-L\leq x_j \leq \tfrac{N}{j}+L$.
%\begin{align}\label{xj}
%  \tfrac{N}{j}-L\leq x_j \leq \tfrac{N}{j}+L
%\end{align}
Since $L$ and $M$ are constants independent of $N$, $x_j=O(\tfrac{N}{j})$ . So we have
\begin{align}
  {x_j}^M-(\tfrac{N}{j})^M &= (x_j-\tfrac{N}{j})\left( {x_j}^{M-1}+{x_j}^{M-2}(\tfrac{N}{j})+\cdots+(\tfrac{N}{j})^{M-1}\right)\nonumber\\
  &=O\left( (\tfrac{N}{j})^{M-1} \right)
\end{align}
Applying this growth estimate to $Z_1$ ,we have
\begin{align}
  Z_1=\sum_{j=1}^{N_0}\mu(j)(\tfrac{N}{j})^M + O\Big( \sum_{j=1}^{N_0}(\tfrac{N}{j})^{M-1} \Big)
\end{align}

Now we turn to $Z_2$. Clearly, if $j$ is larger than $N_0$, there is no more than one integer in $\mathbb{N}$ that can be divided by j. Thus

\begin{align}
    Z_{2} = \sum_{j=N_0+1}^{\infty}\mu(j){x_j}^M %\nonumber \\
    =\sum_{n\in\mathbb{N}}\sum_{j|n \atop j\geq N_0+1}\mu(j)
    = \sum_{n\in \mathbb{N}} \Bigg(\sum_{j|n \atop  j<\infty}\mu(j)- \sum_{j|n \atop  j\leq N_0} \mu(j) \Bigg)
\end{align}
Where $j|n$ means $n$ can be divided by $j$. Based on the property of M\"{o}bius function that
\begin{equation}
    \left\{\begin{array}{cc}
     \sum\limits_{j|n\;and\;j<\infty}\mu(j)=0, & \;\;\;\;\;\;if\,n\neq 1 \\
     \sum\limits_{j|n\;and\;j<\infty}\mu(j)=1, & \;\;\;\;\;\;if\,n=1
   \end{array}\right.
\end{equation}
We have,
\begin{align}
  Z_2=O\Bigg(\sum_{n\in\mathbb{N}}\sum_{j|n \atop j\leq N_0}\mu(j)\Bigg)=O(N^2)
\end{align}
So
\begin{align}\label{P1}
  P=\frac{Z_1+Z_2}{N^M}=\sum_{j=1}^{N_0}\mu(j)\frac{1}{j^M} + O\Big( N^{-1}\sum_{j=1}^{N_0}\frac{1}{j^{M-1}} \Big) +O(N^{2-M})
\end{align}

For the first sum in Equation (\ref{P1}), we rewrite it as
\begin{align}
 \sum_{j=1}^{N_0}\tfrac{\mu(j)}{j^M}=\tfrac{1}{\zeta(M)}- \sum_{j=N_0+1}^{\infty}\tfrac{\mu(j)}{j^M}&=\tfrac{1}{\zeta(M)}+O\left( \int_N^\infty \tfrac{dx}{x^M} \right) \nonumber\\
 &= \tfrac{1}{\zeta(M)}+O(N^{1-M})
\end{align}
where $\zeta(\cdot)$ is Riemann zeta-function.

Since
\begin{align}
  \sum_{j=1}^{N_0}\frac{1}{j^{M-1}}=O&\Big (\sum_{j=1}^{N_0}\frac{1}{j^{M-1}}\Big)=O\Big(1+\int_1^{N_0} \frac{dx}{x^{M-1}} \Big)=O(1)
\end{align}
so the second sum in Equation (\ref{P1}) is $O(N^{-1})$.

Hence, we find that
\begin{align}\label{PO}
  P=\frac{1}{\zeta(M)}+O(N^{-1})
\end{align}
Since the available bandwidth is usually up to dozens of megahertz and $f_{min}$ is very small, $N$ is a very large integer. This makes the second term of (\ref{PO}) a trivial tail, so
\begin{align}\label{P}
  P\approx \frac{1}{\zeta(M)}
\end{align}

\section{Simulation results}
The specification of the PRSHF we deal with in simulation is as follows. Total bandwidth of the transceiver ranges from $54MHz$ to $862MHz$. The minimum resolvable frequency interval is $f_{min}=1kHz$. The number of available frequencies is $N = 2^{15}$. We examine three different scenarios where the number of frequency segments $L$ is 1, 7, and 12 respectively. $M$ ranges from 3 to 13. $M$ frequencies of measurement are randomly chosen from the available frequencies with an m-sequence generator.

Theoretical and simulation results about the probability of $\{ k_i\}$ being relatively prime are given in Fig.1. As can be seen from the figure, simulation results agree with the theoretical one very well in all the three scenarios with respect to the number $L$ of frequency segments.

%A noteworthy fact about the theoretical result given in (\ref{P}) is that it is still valid when $M$ is 2 according to simulation results.

\begin{figure}[h]
\centering{\includegraphics[width=100mm]{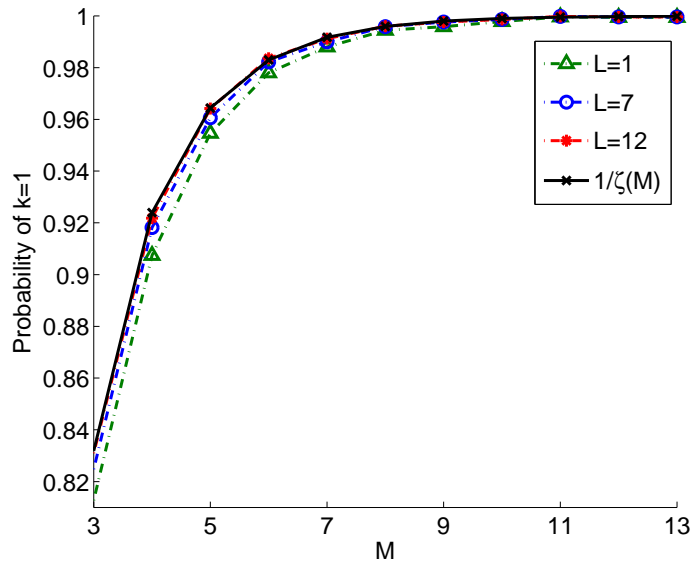}}
\caption{The probability of $M$ positive integers $\{ k_i\}$ chosen from $L$ segments being relatively prime}
\end{figure}

\section{Conclusion}
Approximation (\ref{P}) is a powerful generalization to the result in \cite{6,7} with important implications in the characterization of a PRSHF's unambiguous distance. The most remarkable thing about the result is that if the total number of frequencies $N$ is large enough, the probability that UD in a PRSHF takes on the maximum value depends only on the number $M$ of frequencies employed in measurement, \emph{no matter what the parameter $L$ is or what the distribution of available bandwidth is}.

Another noteworthy consequence of Approximation (14) is that $P>0.999$ when $M>10$.  This means that the UD in a PRSHF almost certainly takes on the maximum value when the number of frequencies employed is greater than 10. Considering that a PRSHF is usually able to cover as vast an area as needed with the maximum value of UD, it should be a highly scalable ranging system with respect to unambiguous distance.

%\section{Conclusion}
%The technical contribution of this letter is best summarized by the simple and elegant expression given in Approximation (\ref{13}). Apart from its theoretical implications, it shows that the probability of UD in a PRSHF takes on the maximum value is almost 1 when the number of frequencies employed in measurement is greater than 10. Since a PRSHF is usually able to cover as vast an area as needed with the maximum value of UD, that means a PRSHF can operate well in large areas effortlessly only if it employs more than 10 randomly chosen frequencies in measurement. To put it another way, a PRSHF is a highly scalable ranging system with respect to unambiguous distance.

\section*{Acknowledgment}
This work was supported by the National Science Foundation of China (61273047 and 61071115)

\end{document}